\documentstyle[aps,prb,multicol,epsf]{revtex}
\begin{document}

\title{
Low-temperature anomaly at the edge of the Heisenberg spin chains:
a boundary conformal field theory approach
}

\author{Satoshi Fujimoto}
\address{
Department of Physics,
Kyoto University, Kyoto 606, Japan
}

\date{\today}
\maketitle
\begin{abstract}
Asymptotically exact low temperature expansions for the $s=1/2$
Heisenberg XXZ chains with boundaries are implemented 
by using the boundary conformal field theory.
It is found that for $1/2<\Delta\leq 1$, ($\Delta$, an anisotropic parameter)
the boundary terms of 
the spin susceptibility and the specific heat coefficient 
show divergent behaviors, as temperature
decreases. The degrees of the divergence coincide with those
obtained by the Bethe ansatz method at zero temperature.
Such low-temperature anomalous behaviors at boundaries
are deeply related with finite-temperature 
corrections of the boundary entropy or the ground state degeneracy, 
the presence of which yields
a very sensitive response of spin excitations.
\end{abstract}

%\begin{multicols}{2}

\section{Introduction}

In the last decade, effects of open boundaries 
in quantum one-dimensional (1D) spin systems have been
extensively studied in connection with impurity 
problems.\cite{al,sk,ve,gri,ji,ka,egg,egg2,de,rom,qin,bru}
From theoretical point of view, advanced
non-perturbative techniques such as 
the Bethe ansatz exact solutions, and 
the boundary conformal field theory have been successfully applied
to the investigation of boundary effects.
\cite{al,sk,ve,gri,ji,ka,egg,egg2,de,rom,qin,bru,fu,ess,as,fr,yam,shi,degu,fuji2}
Although a lot of interesting features of boundary critical phenomena
associated with open ends have been elucidated so far, 
there remain some issues which are not yet fully understood.
One of them is a finite-temperature effect on the boundary 
spin susceptibility.
According to the Bethe ansatz solutions for the spin-1/2
Heisenberg chains, the SUSY $t$-$J$ model, and the Hubbard model,
the presence of boundaries gives rise to
enhanced spin correlation in the vicinity of 
boundaries.\cite{ess,as,fuji2} 
It was found that the boundary part of 
the uniform spin susceptibility at zero temperature
for these systems behaves like $\sim 1/[h\{\ln(h)\}^2]$ at the isotropic
point. Here $h$ is a magnetic field.
This divergent behavior for a small magnetic field implies that
near the boundary spin excitations are very sensitive to
a perturbation induced by external fields or finite temperatures.
It is expected that a similar divergent behavior may also appear in
the temperature dependence of the boundary spin susceptibility.
As was shown in \cite{ess,as,fuji2}, 
the singular contribution at zero temperature
stems from the surface energy perturbed by leading irrelevant
interactions.
It was suggested in ref.\cite{fuji2} that at finite temperatures,
in addition to the surface energy, 
a boundary entropy perturbed by irrelevant interactions also gives rise to
the singular contribution.
The boundary entropy is a residual entropy at zero temperature
caused by the presence of open boundaries.\cite{lud}
In the case that the low-energy fixed point is 
the Tomonaga-Luttinger liquid, namely, the Gaussian theory with
the central charge $c=1$, the boundary entropy is given by,
$S_{\rm B}=\ln(1/\sqrt{2R})$ for
the Dirichlet boundary condition. Here $R$ is a radius parameter
of the Gaussian model.
Thus at finite temperatures or with a magnetic field,
the presence of irrelevant interactions 
yields corrections of $R$ and $S_{\rm B}$,
leading the singular temperature or field dependence of boundary quantities.
 
Unfortunately, because of a technical problem inherent
in boundary integrable systems, it is difficult to
calculate boundary contributions at finite temperatures
in terms of the Bethe ansatz method.
However, we can compute finite-temperature corrections 
by using field theoretical methods 
that is based upon a low energy effective theory. 
For this purpose, the boundary conformal field theory is a very powerful
tool.\cite{cft,car2,car1} 
In this paper, utilizing these techniques, we develop
asymptotically exact low temperature expansions for
the spin-1/2 Heisenberg XXZ chains with open boundaries,
and calculate the leading temperature dependences of
the boundary spin susceptibility and the specific heat coefficient.
It is found that, as expected from the results obtained by the Bethe ansatz 
method, for $1/2<\Delta \leq 1$ ($\Delta$, anisotropic parameter),
these quantities show divergent behaviors, as temperature decreases.
It is expected that the results may be relevant to experimental
observations in real quasi 1D spin systems with non-magnetic impurities.

The organization of this paper is as follows.
In the next section, we briefly review the low energy effective field
theory for the spin-1/2 XXZ chain, and develop
a perturbation theory in the case with open boundaries.
In Sec.III, our main results for the low-temperature dependence of
the spin susceptibility and the specific heat coefficient caused by
boundary effects are presented.
Discussion and summary is given in Sec.IV. 

\section{Low energy effective field theory}

In this section, to establish notations, we first review   
the low energy effective field theory for the $s=1/2$ Heisenberg 
XXZ chains briefly.\cite{luk}
On the basis of this effective theory, we will perform perturbative
expansions in terms of irrelevant interactions 
for the system with boundaries.
The Hamiltonian for the spin-1/2 Heisenberg spin chains is given by,
\begin{equation}
H_{XXZ}=J\sum [S_{i}^xS_{i+1}^x+S_{i}^yS_{i+1}^y+\Delta S_{i}^zS_{i+1}^z].
\label{xxz}
\end{equation}
Here we consider only the antiferromagnetic case $J>0$.
In the massless region, $0\leq\Delta \leq 1$, the low energy fixed point
of (\ref{xxz}) is the Tomonaga-Luttinger liquid, which belongs to
the universality class of the Gaussian theory with the central charge $c=1$.
In this case, 
the low energy effective Hamiltonian with leading irrelevant interactions
is exactly obtained by Lukyanov.\cite{luk}
For $1/2<\Delta<1$, it is written as,
\begin{eqnarray}
H&=&H_0+H_{int}, \label{eff1}\\
H_0&=&\int_0^L \frac{dx}{2\pi}[(\partial_x\phi)^2+\Pi^2], \\
H_{int}&=&a^{2K-2}\lambda\int_0^L \frac{dx}{2\pi}\cos(\sqrt{8K}\phi).
\label{int1}
\end{eqnarray} 
Here $L$ is the linear system size, and the constants $K$, $a$, and 
$\lambda$ are parametrized as,
\begin{equation}
K=[1-\frac{1}{\pi}\cos^{-1}(\Delta)]^{-1},
\end{equation}
\begin{equation}
a=\frac{2(K-1)}{JK\sin(\pi/K)},
\end{equation}
\begin{equation}
\lambda=\frac{4\Gamma(K)}{\Gamma(1-K)}
\left[\frac{\Gamma(1+1/(2K-2))}{2\sqrt{\pi}
\Gamma(1+K/(2K-2))}\right]^{2K-2}.
\end{equation}
The boson fields $\phi(x)$ and $\Pi(x)$ satisfy the canonical conjugate
relation, $[\phi(x),\Pi(x')]=i\pi\delta(x-x')$.
It is convenient to introduce the mode expansion form of $\phi(x)$,
and $\theta(x)\equiv \int^x dx' \Pi(x')$,
\begin{eqnarray}
\phi(x,t)=Q+\frac{\pi}{L}Pt+\frac{2\pi}{L}\frac{Nx}{\sqrt{2K}}
+\frac{i}{2}\sum_{n\neq 0}\frac{1}{n}
(\alpha_n e^{-i\frac{2\pi n}{L}(t+x)}+\bar{\alpha}_n
e^{-i\frac{2\pi n}{L}(t-x)}), \label{mode}
\end{eqnarray} 
\begin{eqnarray}
\theta(x,t)=\tilde{Q}+\frac{\pi}{L}Px+\frac{2\pi}{L}\frac{Nt}{\sqrt{2K}}
+\frac{i}{2}\sum_{n\neq 0}\frac{1}{n}
(\alpha_n e^{-i\frac{2\pi n}{L}(t+x)}-\bar{\alpha}_n
e^{-i\frac{2\pi n}{L}(t-x)}), \label{mode2}
\end{eqnarray}
where the operators satisfy the following commutation relations,
\begin{eqnarray}
[\alpha_n,\alpha_m]=[\bar{\alpha}_n,\bar{\alpha}_m]=n\delta_{n+m,0},
\end{eqnarray}
\begin{eqnarray}
[\alpha_n,\bar{\alpha}_m]=0, \qquad [Q,P]=[\tilde{Q},P]=i.
\end{eqnarray}
$N$ in the right-hand side of (\ref{mode}) is an integer corresponding to
the winding number of the phase field $\phi$.

As will be seen later, boundary states are constructed from 
the highest weight state of the U(1) Kac-Moody algebra
$|M,N \rangle$, which are the eigen states of 
the zero mode of $\alpha_n$, $\bar{\alpha}_n$ defined as,
\begin{eqnarray}
\alpha_0=\frac{P}{2}+\frac{N}{\sqrt{2K}}, \qquad
\bar{\alpha}_0=\frac{P}{2}-\frac{N}{\sqrt{2K}}.
\end{eqnarray}
$| M,N\rangle$ satisfies the eigen value equations,
\begin{eqnarray}
\alpha_0|M,N\rangle&=&
\biggl(\frac{\sqrt{2K}M}{2}+\frac{N}{\sqrt{2K}}\biggr)
|M,N\rangle,  \\ 
\bar{\alpha}_0|M,N\rangle&=&
\biggl(\frac{\sqrt{2K}M}{2}-\frac{N}{\sqrt{2K}}\biggr)
|M,N\rangle .
\end{eqnarray}
The primary field corresponding to $|M,N \rangle$ 
has the conformal dimension,
\begin{equation}
\Delta_{MN}+\bar{\Delta}_{MN}=\frac{1}{2}\bigl(KM^2+\frac{N^2}{K}\bigr).
\label{dim}
\end{equation}

In the case of $0\leq \Delta \leq 1/2$ ($K\geq3/2$), 
the low-temperature anomalous
behaviors at boundaries do not appear, as is easily seen from
the dimensional analysis.
Thus we will not consider this case in the following.

At the isotropic point $\Delta=1$ ($K=1$), the effective low energy theory is
described by the level $k=1$ SU(2) Wess-Zumino-Witten model 
with a marginally irrelevant interaction:
\begin{eqnarray}
H&=&H_{WZW}+H_m, \label{eff2}\\
H_m&=&-g\int_0^L\frac{dx}{2\pi}\sum_{a=1}^3J^a(x)\bar{J}^a(x).\label{int2}
\end{eqnarray}
Here $H_{WZW}$ is the Hamiltonian of the level $k=1$ SU(2)
Wess-Zumino-Witten model, and $J^a(x)$ ($\bar{J}^a(x)$)
is the left (right) moving current of the level $k=1$ SU(2) 
Kac-Moody algebra.
The running coupling constant $g$ depends on
temperature $T$ and an external magnetic field $h$ through the scaling 
equation,\cite{luk}
\begin{eqnarray}
g^{-1}+\frac{1}{2}\ln(g)=-{\rm Re}[\psi(1+\frac{i h}{2\pi T})]
+\ln(\sqrt{2\pi} e^{1/4}J/T), \label{sca}
\end{eqnarray}
with $\psi(x)$ the di-gamma function.

Generally, in systems with boundaries, there may be boundary operators
in addition to bulk interactions.
However, as was pointed out in ref.\cite{bru}, in the absence of
symmetry-breaking external fields at boundaries, 
we can exclude this possibility for the Heisenberg XXZ chains.

In the following, we will implement a perturbative expansion 
of the free energy
in terms of leading irrelevant interactions, 
and evaluate $1/L$-corrections characterizing interesting boundary effects.
For this purpose, following the idea of Cardy and Lewellen,\cite{car1}
we consider the geometry of a semi-infinite cylinder with perimeter $1/T$.
Interchanging space and time coordinates, 
we define the phase field on this geometry,
\begin{eqnarray}
\phi^c(x,t)=Q+\pi TPx+2\pi T\frac{Nt}{\sqrt{2K}}
+\frac{i}{2}\sum_{n\neq 0}\frac{1}{n}
(\alpha_n e^{-i2\pi Tn(x+t)}+\bar{\alpha}_n
e^{-i2\pi Tn(x-t)}). \label{mode3}
\end{eqnarray}
Then, the Hamiltonian on the semi-infinite cylinder
is written as,
\begin{eqnarray}
H^c&=&H^c_0+H^c_{int}, \\
H^c_0&=&
\int^{1/T}_0\frac{dt}{2\pi}[(\partial_t\phi^c)^2+(\partial_x\phi^c)^2], \\
H^c_{int}&=&a^{2K-2}\lambda\int^{1/T}_0\frac{dt}{2\pi}\cos(\sqrt{8K}\phi).
\end{eqnarray}
We express the partition function by using the transfer matrix 
$\exp(-LH^c)$
and the boundary state $|B\rangle$.
The lowest order terms of the free energy for (\ref{eff1}) are given by,
\begin{eqnarray}
F=-\frac{aT}{L}\ln\langle 0|e^{-LH^c_0}|B\rangle
+\frac{aT}{L}\int^L_0 dx\frac{\langle0|\exp(-LH^c_0)
\exp(xH^c_0)H^c_{int}\exp(-xH^c_0)|B\rangle}{\langle0|\exp(-LH^c_0)|B\rangle}
+\cdot\cdot\cdot,
\label{free}
\end{eqnarray}
where $|0\rangle$ is the ground state of $H^c_0$.
The first term of the right-hand side of (\ref{free}) is 
the free energy of the $c=1$ Gaussian model. 
The second term is the $1/L$ correction that
emerges as a result of boundary effects.
Using Cardy and Lewellen's method,\cite{car1} we can compute this term as,
\begin{eqnarray}
\frac{\lambda}{L}\frac{(\pi aT)^{y_{\Phi}}}{2\pi a}
\frac{\langle\Phi|B\rangle}{\langle 0|B\rangle}
\int^L_0dx\frac{1}{[\sinh(2\pi Tx)]^{y_{\Phi}}}, \label{bound}
\end{eqnarray}
where 
$|\Phi\rangle$ is the primary state that corresponds to
the conformal field $\exp(i\sqrt{8K}\phi)$. 
$y_{\Phi}$ is its conformal dimension.
For the isotropic case, a similar consideration is applicable.
In the following, we evaluate (\ref{bound}) exactly for
particular boundary conditions.

\section{Finite-temperature effects on boundary critical phenomena}

\subsection{The anisotropic case $1/2<\Delta<1$}

In the evaluation of the boundary part of the free energy (\ref{free}),
we utilize properties of the boundary state.
A conformally invariant boundary condition is
imposed by demanding $T(z)=\bar{T}(\bar{z})$ at the boundary.\cite{car2}
Here $T(z)$ ($\bar{T}(\bar{z})$) is the holomorophic (anti-holomorophic) part
of the stress energy tensor.  
For the Gaussian model with $c=1$, this condition leads
the following constraint on the boundary state,\cite{ca1,ca2,won,osh} 
\begin{eqnarray}
(\alpha_n\pm \bar{\alpha}_{-n})|B\rangle=0, \label{bc}
\end{eqnarray}
where the plus (minus) sign corresponds to the Neumann 
(Dirichlet) boundary condition.
The solution for (\ref{bc}) is expressed in terms of the 
Ishibashi state.\cite{ish,car1}
For the Dirichlet condition, it is given by,\cite{ca1,ca2,won,osh}
\begin{eqnarray}
|D\rangle =\biggl(\frac{K}{2}\biggr)^{1/4}\sum_{M=-\infty}^{\infty}
e^{-i\sqrt{2 K}M\phi_0}
\exp\biggl(-\sum_{n=1}^{\infty}\frac{\alpha_{-n}\bar{\alpha}_{-n}}{n}\biggr)
|M,0 \rangle .  \label{dir}
\end{eqnarray}
Note that the prefactor $(K/2)^{1/4}$ is the boundary entropy 
at zero temperature.
The constant $\phi_0$ is the eigen value of the boson field
$\phi$ at the boundary. 
The boundary state for the Neumann condition is,\cite{ca1,ca2,won,osh}
\begin{eqnarray}
|N\rangle =\biggl(\frac{1}{2K}\biggr)^{1/4}\sum_{N=-\infty}^{\infty}
e^{-i\sqrt{2 /K}N\theta_0}
\exp\biggl(\sum_{n=1}^{\infty}\frac{\alpha_{-n}\bar{\alpha}_{-n}}{n}\biggr)
|0,N \rangle , \label{neu}
\end{eqnarray}
where $\theta_0$ is the boundary value of the dual field, 
$\theta(x)$.
The prefactor $(1/2K)^{1/4}$ is the boundary entropy for this 
boundary condition.

The matrix element between $|0\rangle$ and $|B\rangle$ appeared in
(\ref{bound}) can be computed by using (\ref{dir}) and (\ref{neu}).
For the case of $1/2<\Delta<1$ ($1<K<3/2$), the leading irrelevant interaction
(\ref{int1}) is expressed in terms of 
the primary field $\exp(i\sqrt{8K}\phi)$ which has the conformal dimension
$2K$. From (\ref{dim}), we see that $|\Phi\rangle$ 
is the primary state $|2,0\rangle$.
$\langle \Phi |B\rangle$ is non-vanishing, only if $|B\rangle$
contains $|2,0\rangle$. The Neumann boundary state (\ref{neu}) does
not satisfy this condition, leading $\langle \Phi |N\rangle =0$.
On the other hand, $\langle \Phi |D\rangle$ gives a finite contribution.
In the following, we put $\phi_0=0$.
This choice of the averaged value of the phase at the boundary
is consistent with the free open boundary condition for 
the boundary magnetization.
%This choice of the averaged value of the phase at the boundary
%does not affect boundary quantities.
Then, we have,
\begin{eqnarray}
\frac{\langle 2,0 |D\rangle}{\langle 0| D\rangle}=1. \label{prefac}
\end{eqnarray}

In the presence of an external magnetic field $h$, 
we add the Zeeman energy term,
\begin{eqnarray}
H_Z=-S_zh=-\frac{h}{\pi}\sqrt{\frac{K}{2}}\int^L_0 dx \partial_x\phi, 
\end{eqnarray} 
to the Hamiltonian (\ref{eff1}).
The free energy is evaluated by shifting the boson field $\phi(x)$
to $\tilde{\phi}(x)=\phi(x)-\sqrt{K/2}hx$.
Then the integral in (\ref{bound}) is replaced by,
\begin{eqnarray}
\int^L_0dx\frac{\cos(2Khx)}{[\sinh(2\pi Tx)]^{y_{\Phi}}}. \label{integ}
\end{eqnarray}
Combining (\ref{prefac}) and (\ref{integ}), and carrying out
the integral, we obtain corrections to the boundary part of the free energy,
\begin{eqnarray}
\delta F_{\rm B}=-\frac{\lambda}{4\pi L}(2\pi a T)^{2K-1}{\rm Re}
[B(K+i\frac{Kh}{2\pi T},1-2K)], \label{freeb}
\end{eqnarray}
where $B(x,y)=\Gamma(x)\Gamma(y)/\Gamma(x+y)$.

Using (\ref{freeb}),  we can calculate the boundary contribution to
the spin susceptibility,
\begin{eqnarray}
\chi_{\rm B}&=&-\biggl.\frac{\partial^2 \delta F_{\rm B}}{\partial h^2}
\biggr|_{h=0} \nonumber \\
&=&\frac{\lambda}{L}\frac{a^2K^2}{4\pi}
B(K,1-2K)[\pi^2-2\psi'(K)](2\pi a T)^{2K-3}, \label{chi1}
\end{eqnarray}
where $\psi'(x)=d\psi(x)/dx$. 
Note that for $1<K<3/2$ ($1/2<\Delta<1$), the boundary spin
susceptibility $\chi_{\rm B}$ shows a divergent behavior $\sim 1/T^{3-2K}$,
as temperature decreases.
This anomalous temperature dependence is also observed in the boundary part 
of the specific heat coefficient computed as,
\begin{eqnarray}
\frac{C_{\rm B}}{T}&=&-\frac{\partial^2 \delta F_{\rm B}}{\partial T^2} 
\nonumber \\
&=&\frac{\pi a^2\lambda}{L}(2K-1)(2K-2)B(K,1-2K)(2\pi a T)^{2K-3}.
\label{heat1}
\end{eqnarray}
In the above derivation, we have neglected the boundary terms
of the low energy fixed point which are regular in $h$ and $T$.
We would like to stress that in the formulas (\ref{chi1}) and (\ref{heat1})
there is no free parameter, and the prefactors are exactly obtained.
These divergent behaviors are physically understood as follows.
In contrast to the bulk Heisenberg chains in which
the ground state is a spin singlet state, 
spin singlet formation in the vicinity of boundaries is strongly
disturbed by thermal fluctuation, because of the enhanced correlation
at the boundaries which stems from the ground state degeneracy.
It should be emphasized that
the singular behaviors are not due to the presence of boundary operators,
but interpreted as a consequence of finite-temperature corrections of
the surface energy and the boundary entropy $\ln\langle 0|B\rangle$ 
caused by irrelevant interactions.

At zero temperature with a finite magnetic field,
a similar singular behavior appears in the field dependence of
the boundary spin susceptibility given by,
\begin{eqnarray}
\chi_{\rm B}(T=0)=-\frac{\lambda}{L}\frac{(aK)^{2K-1}}{4\pi}(2K-1)
\Gamma(1-2K)h^{2K-3}. \label{zerospin}
\end{eqnarray}
The zero temperature susceptibility is also derived from
the Bethe ansatz exact solution by using the Wiener-Hopf method.
We have checked that (\ref{zerospin}) coincides with the result obtained
by the Bethe ansatz. 

\subsection{The isotropic case $\Delta=1$}

In the isotropic case $K=1$, let us consider
the boundary condition that preserves the SU(2) spin rotational symmetry,
$J^a(x=0)=\bar{J}^a(x=0)$, which leads,\cite{car2}
\begin{eqnarray}
(J^a_n+\bar{J}^a_{-n})|B\rangle =0,
\label{bo2}
\end{eqnarray}
for $a=1,2,3$.
Here $J^a_n$ ($\bar{J}^a_n$) are the $n$-th mode component
of the left (right) moving SU(2) current.
The representation of $|B\rangle$ 
for this boundary condition was precisely discussed 
by Gaberdiel and Gannon.\cite{gg}
However, instead of using the explicit formula of $|B\rangle$,
we exploit the SU(2) current algebra to evaluate the prefactor
of (\ref{bound}).
The marginal interaction (\ref{int2}) consists of
the first descendants in the Verma module of the identity operator.
The corresponding state vector is 
$|\Phi\rangle=J^a_{-1}\bar{J}^a_{-1}|0\rangle$.
Using the commutation relations 
for the level-1 SU(2) Kac-Moody algebra,\cite{kac}
\begin{eqnarray}
[J^a_n,J^b_m]=i\epsilon_{abc}J^c_{n+m}+\frac{n}{2}\delta_{n+m,0}\delta_{ab},
\end{eqnarray}
and $\langle 0|J_{-1}^a=0$, we have,
\begin{eqnarray}
\sum_{a=1}^3\frac{\langle 0|J^a_1\bar{J}^a_1 |B\rangle}
{\langle 0 |B\rangle}=
-\sum_{a=1}^3\frac{\langle 0|J^a_1J^a_{-1}|B\rangle}{\langle 0 |B\rangle}=
-\frac{3}{2}. 
\end{eqnarray}
In the evaluation of the integral in (\ref{bound}), we circumvent
unphysical divergence by taking the limit, 
\begin{equation}
\lim_{K\rightarrow 1}B(K,1-2K)=-1.
\end{equation}
Finally, we end up with,
\begin{eqnarray}
\delta F_{\rm B}=-\frac{3aT}{4L}g. \label{free2}
\end{eqnarray}
In the above derivation, we take into account
effects of an external magnetic field only through the
field dependence of the coupling constant $g$, because
it gives leading corrections induced by the magnetic field.
Using (\ref{sca}) and (\ref{free2}), we obtain 
the leading term of the boundary spin susceptibility and
the specific heat coefficient,
\begin{eqnarray}
\chi_{\rm B}=-\frac{2a\psi''(1)}{16\pi^2 L}
\frac{1}{T[\ln(\alpha J/T)]^2}, \label{chi2}
\end{eqnarray}
\begin{eqnarray}
\frac{C_{\rm B}}{T}=\frac{3a}{4L}\frac{1}{T[\ln(\alpha J/T)]^2},
\label{spe2}
\end{eqnarray}
where $\psi''(1)=-2.40411...$, and $\alpha=\sqrt{2\pi}\exp(1/4+\gamma)$
with $\gamma$ the Euler constant.
The temperature dependence of these boundary quantities 
is analogous to the field dependence at zero temperature,
$\sim 1/h[\ln(h)]^2$, which is  
obtained from the Bethe ansatz solution for the supersymmetric $t$-$J$ model
and the Hubbard model.\cite{ess,as,fuji2}
Because of the logarithmic factor, the divergent behaviors of
(\ref{chi2}) and (\ref{spe2}) are observable only at very low temperatures
$T\ll \alpha J\sim 5.73251 J$.
Note that the expression (\ref{free2}) is applicable for $h\ll T$.
Thus we cannot derive the zero temperature susceptibility from (\ref{free2}).
We need to compute higher order corrections in terms of $h$ to
take the limit $T\rightarrow 0$.
This task is not straightforward because the presence of
a term like $\sim |h|/[\ln(h)]^2$ in the free energy implies that
the perturbative expansion in $h$ is singular.
It is required to take the limit $K\rightarrow 1$
from the anisotropic point carefully,
as was done for the bulk system by Lukyanov.\cite{luk}

A logarithmic singularity was also found in the NMR relaxation rate
at the edge of the system calculated by Affleck and Qin, and Brunel et al., 
who showed that the bulk marginal interaction
gives rise to an anomalous dimension at boundaries.\cite{qin,bru}
The above logarithmic singularity has the same origin as that for
the NMR relaxation rate at boundaries.

\section{Discussion and Summary}

The Curie-like temperature dependence with logarithmic corrections
of the boundary spin susceptibility (\ref{chi2}) 
may be relevant to experimental observations.
According to experimental measurements of the spin susceptibility for
the Heisenberg spin chains such as ${\rm Sr_2CuO_3}$,
a Curie-like behavior at low temperature region is always
observed, but has been regarded as an extrinsic impurity effect.\cite{uchi}
However, the results obtained in this paper imply that
such an effect may be intrinsically important for the Heisenberg spin chains
with an open end.
In particular, it probes the character of leading irrelevant interactions
in the bulk.
To observe the effects obtained in this paper experimentally,
more systematic studies distinguishing intrinsic and extrinsic effects
are required.

Although the boundary spin susceptibility at finite temperatures
was studied before by both numerical and field theoretical methods,
the anomalous temperature dependence has not been 
found so far.\cite{egg,egg2,rom}
Actually, the divergent behaviors of the boundary contributions
(\ref{chi2}) and (\ref{spe2}) are visible only 
at very low temperatures, because of the logarithmic reduction
factor and the small magnitude of the prefactors.
It seems that these equations do not contradict with the numerical results
obtained in refs. \cite{egg2} and \cite{rom}.
In the previous works based on the field theoretical methods\cite{egg,egg2}, 
effects of irrelevant interactions on
the surface energy and the boundary entropy, which are the most
important ingredient for the divergent behaviors,
have not been calculated explicitly.

%because that, 
%in the previous field theoretical calculations, 
%contributions from the boundary entropy perturbed by irrelevant 
%interactions were not taken into account.
The results obtained here imply that
the temperature dependences of boundary quantities 
similar to (\ref{chi2}) and (\ref{spe2}) should be also found in
the 1D Hubbard model, since its low energy effective theory for 
the spin sector belongs
to the same universality class as that of the spin-1/2 isotropic
Heisenberg chains.
The results derived by the Bethe ansatz method 
at zero temperature for the Hubbard model
is consistent with the present results for the isotropic case, supporting
the presence of anomalous behaviors at boundaries. 

In summary, we have obtained the boundary contributions of the
spin susceptibility and the specific heat coefficient for the spin-1/2
Heisenberg chains using the exact effective low energy theory and
the boundary conformal field theory.
These quantities show divergent behaviors as the temperature is lowered,
indicating strongly enhanced spin correlation 
in the vicinity of the open boundary.

\acknowledgments{}
The author would like to thank S. Eggert, F.H.L. Essler, and 
N. Kawakami for valuable discussions.
This work was partly supported by a Grant-in-Aid from the Ministry
of Education, Science, Sports and Culture, Japan.

%\end{multicols}
%\begin{multicols}{2}
%%%%%
%%%%%%%%%%%%%%%%%%%%%%%%%%%%%%%%%%%%%%%%%%%%%%%%%

%\end{multicols}
                                                                    
%%%%%%%%%

\begin{references}
%%%%%%%%%%%%%%%%%%%%%%%
%%%%%%%%%%%%%%%%%%%%%%%%%%%%%%%%
%\bibitem{ga} M. Gaudin, Phys. Rev. A{\bf 4}, 386 (1971).

\bibitem{al} F.C. Alcaraz, M. N. Barber, M. T. Batchelor,
R.J. Baxter, and G.R.W.Quispel, J. Phys. A{\bf 20}, 6397 (1987).

\bibitem{sk} E. Sklyanin, J. Phys. A{\bf 21}, 2375 (1988).

\bibitem{ve} H. J. de Vega and A. Gonzalez-Ruiz, J. Phys. A{\bf 27},
6129 (1994).

\bibitem{gri} M. T. Gisaru, L. Mezincescu, and R. I. Nepomechie,
J. Phys. A{\bf 28}, 1027 (1995).

\bibitem{ji} M. Jimbo, R. Kedem, T. Kojima, H. Konno, and T. Miwa,
Nucl. Phys. B{\bf 441}, 437 (1995).

\bibitem{ka} A. Kapustin and S. Skorik, J. Phys. A {\bf 29}, 
1629 (1996).

\bibitem{egg} S. Eggert and I. Affleck, Phys. Rev. B{\bf 46}, 10866
(1992).

\bibitem{egg2} S. Eggert and I. Affleck, Phys. Rev. Lett. {\bf 75},
934 (1995).

\bibitem{de} P. de Sa and A. M. Tsvelik, Phys. Rev. B{\bf 52}, 3067
(1995).

\bibitem{rom} S. Rommer and S. Eggert, Phys .Rev. B{\bf 59}, 6301 (1999).

\bibitem{qin} I. Affleck and S. Qin, J. Phys. A{\bf 32}, 7815 (1999).

\bibitem{bru} V. Brunel, M. Bocquet, and Th. Jolicoeur,
Phys. Rev. Lett. {\bf 83}, 2821 (1999).

\bibitem{fu} S. Fujimoto and N. Kawakami, Phys. Rev. B{\bf 54}, 5784
(1996).

\bibitem{ess} F.H.L. Essler, J. Phys. A{\bf 29}, 6183 (1996). 

\bibitem{as} H. Asakawa and M. Suzuki, J. Phys. A{\bf 29}, 225 (1996);
{\it ibid} {\bf 29}, 7811 (1996).

\bibitem{fr} G. Bed\"urftig and H. Frahm, J. Phys. A{\bf 30}, 4139
(1997).

\bibitem{yam} O. Tsuchiya and T. Yamamoto, J. Phys. Soc. Jpn. {\bf 66},
1950 (1997).

\bibitem{shi} M. Shiroishi and M. Wadati, J. Phys. Soc. Jpn. {\bf 66},
1 (1997).

\bibitem{degu} R. Yue and T. Deguchi, J. Phys. A{\bf 30}, 8129 (1997).

\bibitem{fuji2} S. Fujimoto, Phys. Rev. B{\bf 63}, 024406 (2000).

\bibitem{lud} I. Affleck and A. W. W. Ludwig, Phys. Rev. Lett.{\bf 67},
161 (1991).

\bibitem{cft} A. A. Belavin, A. M. Polyakov, and A. B. Zamolodchikov,
Nucl. Phys. B{\bf 241}, 333 (1984).

\bibitem{car2} J. L. Cardy, Nucl. Phys. B{\bf 324}, 581 (1989).

\bibitem{car1} J. L. Cardy and D. C. Lewellen, Phys. Lett. B{\bf 259},
274 (1991).

\bibitem{luk} S. Lukyanov, Nucl. Phys. B{\bf 522}, 533 (1998).

\bibitem{ca1}C. G. Callan, C. Lovelace, and C. R. Nappi, and S. A. Yost,
Nucl. Phys. B{\bf 293}, 83 (1987).

\bibitem{ca2} C. G. Callan, I. R. Klebanov, A. W. W. Ludwig, and
J. M. Maldacena, Nucl. Phys. B{\bf 422}, 417 (1994).

\bibitem{won} E. Wong and I. Affleck, Nucl. Phys. B{\bf 417}, 403 (1994). 

\bibitem{osh} M. Oshikawa and I. Affleck, Nucl. Phys. B{\bf 495}, 533
(1997).

\bibitem{ish} N. Ishibashi, Mod. Phys. Lett. A{\bf 4}, 251 (1989).

\bibitem{gg} M. R. Gaberdiel and Y. Gannon, Nucl. Phys. B{\bf 639}, 
471 (2002).

\bibitem{kac} V. G. Knizhnik and A. B. Zamolodchikov, Nucl. Phys. B{\bf 247},
83 (1984).

\bibitem{uchi} N. Motoyama, H. Eisaki, and S. Uchida, Phys. Rev. Lett.
{\bf 76}, 3212 (1996).
 
\end{references}
\end{document}